\documentclass{aa}
\usepackage{graphicx}
\usepackage{epsfig}
\usepackage{natbib}
\usepackage{txfonts}

\begin{document}
 
\title{The discovery of an expanding X-ray source in the HH\,154 protostellar jet}

\author{F. Favata\inst{1}  \and R. Bonito\inst{2,3} \and G. Micela\inst{2} \and M. Fridlund\inst{1} \and S. Orlando\inst{2} \and S. Sciortino\inst{2} \and G. Peres\inst{3}}

\institute{Astrophysics Missions Division, Research and Scientific Support Department of ESA/ESTEC, Postbus 299, 2200 AG Noordwijk, The Netherlands \and INAF -- Osservatorio Astronomico di Palermo, Piazza del Parlamento 1 90134 Palermo, Italy \and Dipartimento di Scienze Fisiche ed Astronomiche, Sez. Astronomia, Universit\`a di Palermo, Piazza del Parlamento 1, 90134 Palermo, Italy}

\abstract 
   {Protostellar jets are a new class of X-ray sources which has been discovered with both XMM-\emph{Newton} and \emph{Chandra}. The mechanism responsible for the X-ray emission is still not clear. Self-shocking in jets, shocks where the jet hits the surrounding medium, reflected or scattered stellar X-ray emission have all been invoked as possible explanations.}
   {One key diagnostic discriminating among physical emission mechanisms is the motion of the X-ray source: hydrodynamical numerical models of continuous protostellar jets plowing through a uniform medium show an X-ray emitting shock front moving at several hundreds km/s. In the nearest X-ray emitting protostellar jet, HH\,154, this is detectable, with the spatial resolution of the \emph{Chandra} X-ray observatory, over a few years baseline, allowing a robust discrimination among different mechanisms. }
   {We have performed, in October 2005, a deep \emph{Chandra} X-ray observation of HH\,154. Comparison with the previous (2001) \emph{Chandra} observation allows to detect proper motion down to the level predicted by models of X-ray emitting shocks in the jet.}
   {The 2005 \emph{Chandra} observation of HH\,154 shows unexpected morphological changes of the X-ray emission in comparison with the 2001 data. Two components are present: a stronger, point-like component with no detectable motion and a weaker component which has expanded in size by approximately 300 AU over the 4 years time base of the two observations. This expansion corresponds to approximately 500 km/s, very close to the velocity of the X-ray emitting shock in the simple theoretical models.}
  {The 2005 data show a more complex system than initially thought (and modeled), with multiple components with different properties. The observed morphology is possibly indicating a pulsed jet propagating through a non-homogeneous medium, likely with medium density decreasing with distance from the driving source. Detailed theoretical modeling and deeper X-ray observations will be needed to understand the physics of this fascinating class of sources.}


   \maketitle
   
 \section{Introduction}
 
 Soft X-ray emission from jets driven by protostellar sources has been discovered in 1999 thanks to the improved sensitivity and spatial resolution offered by both the XMM-\emph{Newton} and \emph{Chandra} observatories. HH\,2 in Orion has been the first source detected (\citealp{pfg+2001}), with HH\,154 (the jet originating from the embedded binary protostar IRS\,5 in the L1551 star-forming region) following shortly thereafter (\citealp{ffm+2002}; \citealp{bfr2003}). While more sources have been detected later, HH\,154 remains the best source to study thanks to both its being nearby ($d \simeq 150$ pc) and to its being observable through a moderate absorbing column density ($A_V \simeq 7$ mag) while the parent star IRS 5 is hidden by ca. 150 mag of absorption. This allows to study the X-ray emission from the jet with good detail and without any glare from the parent star.

X-ray emission from HH\,154 was observed with XMM-\emph{Newton} (\citealp{ffm+2002}), allowing to determine the temperature of the source at about 4 MK. The limited spatial resolution of XMM-\emph{Newton} did not allow to pinpoint the source of X-ray emission, and \cite{ffm+2002} speculated that the source of X-ray emission might be located within the jet's working surface (i.e.\ the shock formed when the jet hits the circumstellar material), located in 2005 at some 15 arcsec from the parent star. The spatial resolution of \emph{Chandra} allowed \cite{bfr2003} to reject this hypothesis, showing that the X-ray emission is located very close to the protostar itself and thus to the jet's base. 

A number of hypotheses have been advanced relative to the physical mechanism causing the X-ray emission. \cite{bfr2003} speculated on the various possibilities, including reflected or scattered stellar X-ray emission, internal shocks in the jet and shocks formed when the jet hits the cavity wall. Later, \cite{bop+2004} performed detailed hydrodynamic numerical modeling of continuous jets propagating in a uniform medium, showing that the X-ray emission observed from HH\,154 can be explained in such a scenario as coming from the shock forming where the jet interacts with ambient matter, with reasonable assumptions about e.g.\ densities and velocities for both the jet and the medium.
 
Other proposed mechanisms for the origin of the observed X-ray emission predict different properties, which can be used as discriminating diagnostics. For example, scattered or reflected stellar X-ray emission should show the same temporal and spectral variability of the emission from the parent star, in particular the ubiquitous flaring observed in PMS stars. On the other hand, the numerical models of \cite{bop+2004} show the X-ray emitting shocks to be moving at speeds of ca. 500 km/s, with little temporal variability. 

The motion of the X-ray emitting shock predicted by \cite{bop+2004}'s models is sufficiently large as to be detectable with the high spatial resolution of \emph{Chandra} over a temporal baseline of a few years. With the main aim of constraining the physical mechanism driving the X-ray emission from the jet we have performed, in 2005, a deep \emph{Chandra} observation of HH\,154. Its detailed comparison with the 2001 observation allows a strong test of whether the X-ray source indeed displays proper motion, and thus a test of whether the "moving shock" model is indeed consistent with data.
 
    \begin{figure*}[tbh]
   \centering
   \includegraphics[width=12cm]{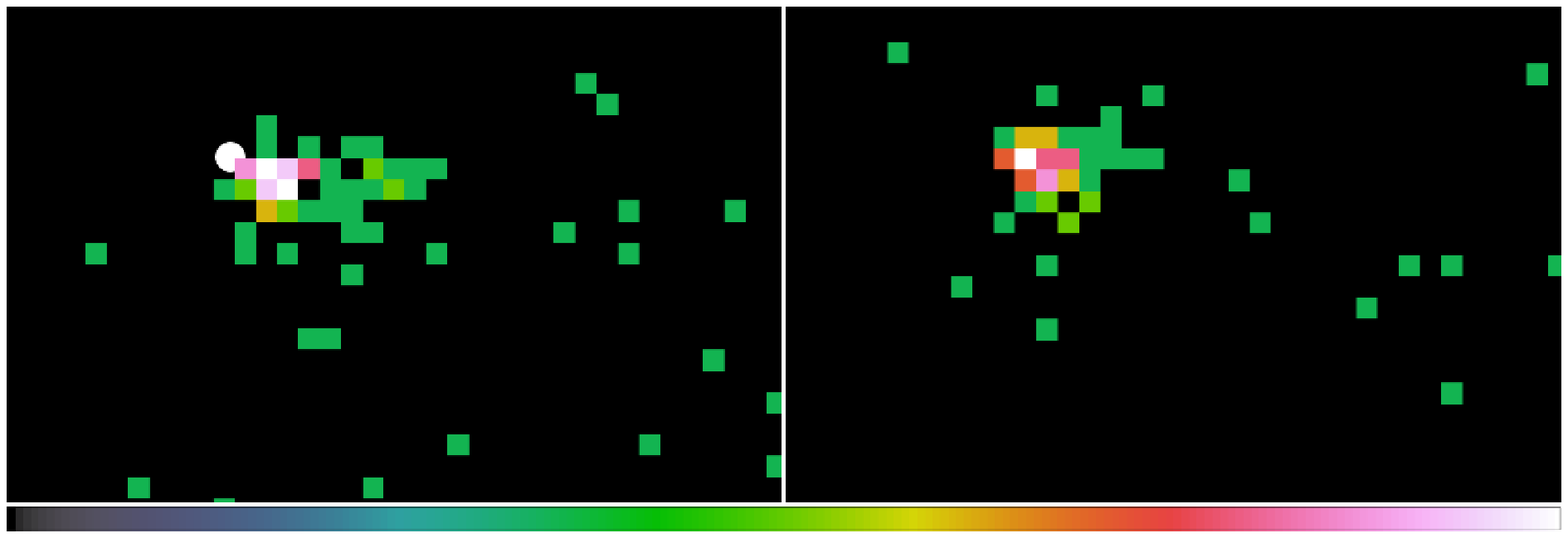}
   \includegraphics[width=12cm]{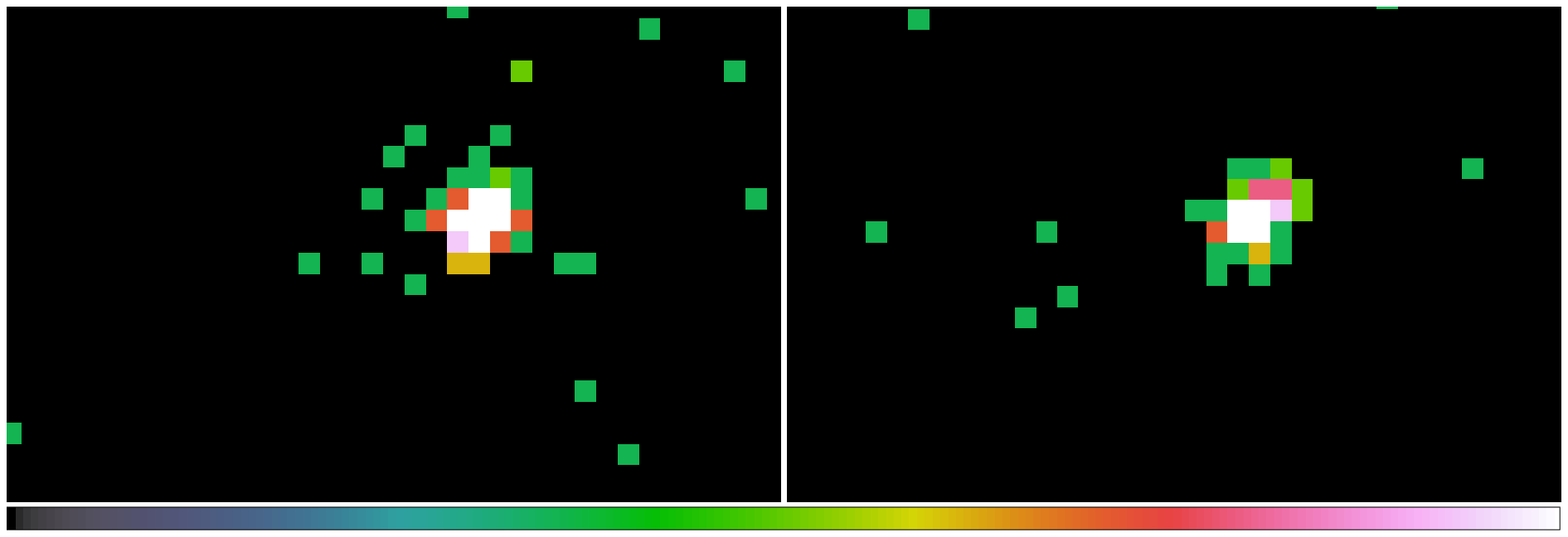}
   \caption{The top panel shows the X-ray source associated with the HH\,154 jet, in the 2005 (left) and 2001 (right) \emph{Chandra} observations. The small white circle in the 2005 observation shows the position of the protostar driving the jet (L1551 IRS5, at 04:31:34.2 $+$18:08:05). The scale of the image is 0.5 arcsec/pixel, north is up, east left. While a hint of an extended component is present already in the 2001 data (as discussed by \citealp{bfr2003}), the 2005 observation clearly shows two components to be present, a point-like source which has the same position as in 2001, and an expanding component which moved, in the jet's direction and away from the driving star, by approximately 2 arcsec between 2001 and 2005. The color scale is the same for both observations. For comparison the bottom panels show (in the same color scale) a background AGN at 2.5 arcmin from HH\,154, demonstrating in both cases the sharpness of the \emph{Chandra} PSF for point-like sources. }
              \label{fig:jets}%
    \end{figure*}

\section{Data analysis and results}
 
 The \emph{Chandra} observation used here was performed on 27 Oct. 2005, for a total duration of 97 ks. The observation was centered at 04:31:32.9, $+$18:08:04.7 (FK5) and was performed with the ACIS-I camera. We processed the observation using the CIAO package, and filtered the data to retain only photons in the 0.3 to 4.0 keV band. To allow an uniform comparison, we have also reprocessed the 2001 observation (described by \cite{bfr2003}) using the same approach. Individual X-ray sources have been detected using the standard wavelet-based algorithm of CIAO. A small positional shift is present between the two observations, as shown by both the stellar and the extragalactic X-ray sources in the field. To allow detection of the expected small proper motion of the jet X-ray source, we have registered the two images by applying a shift determined by minimizing the coordinate differences of the 7 brightest X-ray sources in the field which are also members of L1551 (to minimize the effects of proper motion). The shift between the two observations is $2.43 \pm 0.28$ arcsec, a value within the absolute attitude reconstruction accuracy expected for \emph{Chandra}.
 
 The first surprising result of our 2005 observations is that the X-ray source associated with the jet is clearly extended, as shown in Fig.~\ref{fig:jets}, which shows both the 2001 and the 2005 \emph{Chandra} observations of HH\,154. The position of the driving protostar (L1551 IRS5) is also plotted, from the VLA observations of \cite{rcc+2003}.
 
  Already for the 2001 observation \cite{bfr2003} had suggested that the jet X-ray source might show a hint of being extended; in our 2005 observation not only do we confirm that the source indeed has an extended component, but we also show that the morphology has changed, with the size of the extended component clearly increasing with respect to 2001. The bottom panels of Fig.~\ref{fig:jets} show the image of a background AGN located at 2.5 arcmin from the jet source. In both 2001 and 2005 the source is clearly point-like, and with a different morphology than the jet, showing the latter to be resolved by \emph{Chandra}.
 
\begin{figure}[!h]
   \centering
   \epsfig{width=8cm, clip=, bbllx=20, bblly=150, bburx=570, bbury=410, file=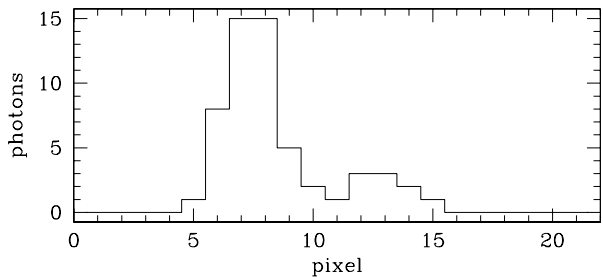}
   \caption{A horizontal cut through the 2005  \emph{Chandra} HH\,154 image, obtained by adding the number of photons in the two rows where the stationary source peaks, showing the relative intensity of the stationary and the moving components. }
              \label{fig:cut}%
    \end{figure}

 In the 2005 observation the source appears to have both an unresolved, point-like component and an extended component. While the unresolved component does not show a detectable motion with respect to 2001, the extended component has clearly become more elongated along the jet's direction, as expected from a shock moving within the jet and away from the parent star.The total count rate from both sources is compatible with their being constant, with no short-term variability visible. Fig.~\ref{fig:cut} shows a horizontal cut through HH\,154 in the 2005 observation, showing the relative intensity of the stationary and moving sources.
 
\begin{figure}[tbh]
   \centering
   \includegraphics[width=8cm]{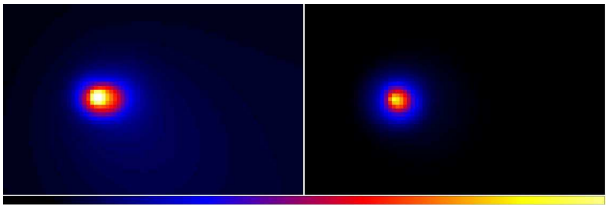}
   \caption{The smoothed image of the X-ray sources associated with HH\,154 are shown, in the left panel for the 2005 data and in the right panel for the 2001 data. The 2005 data show the presence of an elongated component not present in the 2001 data.}
              \label{fig:smooth}%
    \end{figure}

 To confirm the extended nature of the source in the 2005 observation, we performed an adaptive smoothing of the data with a Gaussian kernel, using the software tool \textsc{csmooth} available in the CIAO software. The results, shown in Fig.~\ref{fig:smooth} clearly show the source associated with HH\,154 to be elongated in 2005. In 2001 the same source was compatible with being point-like. At the same time, the AGN as shown in Fig.~\ref{fig:jets} has a symmetric, point-like shape in both smoothed images (not shown in Fig.~\ref{fig:smooth}). This again demonstrates the change in morphology of the X-ray source associated with the jet over the 4 years.
 
 Between the 2001 and 2005 observations the source extent has increased by 2 arcsec, which, at the distance of L1551, correspond to $4.5\times 10^{15}$ cm, or 300 AU. This results in a projected velocity of ca. 330 km/s, which, assuming the inclination of the jet to be 45 deg (\citealp{lfl2005}), implies a shock velocity of ca. 460 km/s, or about 100 AU/yr.

We have also performed a simple spectral analysis of the source's emission. The integrated source has 63 photons, which we have fit with a simple absorbed thermal plasma (\textsc{apec} model in XSPEC), resulting in a temperature of approximately 0.5 keV and a column density of $1.5\times 10^{22}$ cm$^{-2}$. This is essentially identical to the integrated X-ray spectral characteristics determined with the Sep. 2000 XMM-\emph{Newton} observation, and compatible with the \emph{Chandra} 2001 observation, showing limited spectral variability in the source, even in the presence of significant morphological evolution. The number of X-ray photons present in the moving component of the source is too small to attempt a separate spectral analysis. While the energy distribution of the photons from the moving source is compatible with the spectrum of the non-moving component, there is a hint of more photons present at lower energy, which would be compatible either with a lower absorbing column density (i.e.\ with the moving component coming out of the large absorption region in which the jet is born) or with a lower temperature.

The count rate from the integrated source in the 2005 observation is 0.65 cts/ks, compatible within the Poisson errors with the count rate observed in the \emph{Chandra} 2001 observation (0.77 cts/ks), showing limited (if any) changes in the source's luminosity (given the lack of observed spectral evolution).

\section{Modeling implications}

The scenario of \cite{bop+2004}, which the observation discussed here was designed to test, made clear predictions about the evolution of the X-ray emission due to the jet's shock. In particular, in the model which best reproduced HH\,154's X-ray luminosity and spectrum the X-ray source was point-like, located at the head of the jet and had a continuous motion of about 500 km/s. 

The observational results discussed here are more complex than expected: the X-ray source displays both an apparently point-like component, with no measured motion, and a probably extended component which is moving (from the difference in position between the 2001 and 2005 observations) at about 500 km/s, a value very close to the value predicted by the \cite{bop+2004} model. However, the apparently stationary component is not in agreement with the simple \cite{bop+2004} model.

The assumptions behind this model were simple, aiming at elucidating the basic physics. In particular, the jet is assumed to be continuously powered, with constant density of inflowing material, and the ambient medium through which the jet propagates is assumed to be uniform. Neither of these two assumptions is likely to be true in realistic conditions. 

The optical observations of HH\,154 in the last two decades show evidence of the jet being highly inhomogeneous, with density varying by some two orders of magnitude within the jet itself (\citealp{fld+2005}), so that the jet is likely to be driven in an eruptive, rather than continuous, form, with new visible features appearing on a time scale of few years. Evidence for the jet to be pulsed, rather than continuous, comes also from the decreasing velocity of individual blobs through the jet: while the de-projected velocity from optical spectroscopy is up to 600 km/s close to the star, at 3 arcsec from it it begins to drop, decreasing down to 160 km/s for the working surface which is located at some 15 arcsec from the star in the 2005 observations -- see \cite{fld+2005} for details. In the presence of a continuous driving of the jet the shock front would be expected to travel at constant velocity all the way to 15 arcsec; the observed velocity decrease is thus likely indicating a significant pulsating component.

At the same time, the assumption of a homogeneous medium through which the jet propagates is unlikely to be correct: the high level of structuring of the medium is shown for example by the fastly varying circumstellar absorption (the driving star being hidden by some 150 mag of material, while the jet has less than 10 mag of absorption). Even along the jet optical observations show the absorption varying between 2 and at least 6 mag (\citealp{fld+2005}).

Although the predictions of the simple model of a continuous jet ramming into a homogeneous medium of \cite{bop+2004} is not compatible with these observational results, the high level of structuring in both the jet and the medium can possibly explain the phenomenology observed in X-rays, with the same basic physics on which the simple models are based. The density of circumstellar material is very likely to decrease significantly with distance from the jet-driving source (as observed in the structure of the absorbing material), and thus the individual jet pulses (or ``bullets'') will initially meet a dense medium, and then later propagate through a thinner medium. 

\cite{bop+2006} have carried out, building on the approach first used by \cite{bop+2004}, an extensive exploration of the parameter space of both density contrast and velocity for a uniform, constantly driven jet. One result is that the X-ray luminosity of the shock front depends on both the density contrast and the Mach number with which the jet plows through the medium. In the presence of a decreasing ambient density and increasing temperature (due to pressure equilibrium), the jet's luminosity will decrease as it plows though the medium. In addition, in the case of an isolated traveling plasma bullet, the temperature of the X-ray emitting plasma is expected to gradually decrease due to radiative and conductive losses. 

One possible scenario explaining the observed phenomenology is therefore that the driving source is continuously injecting fresh blobs of material into the jet. As each blob rams into the high density medium near the star it will cause the X-ray emission seen as the stationary X-ray source closer to the protostar. As a given blob plows through the high density medium and starts to move through the thinner medium farther away from the star, its X-ray luminosity will decrease while it moves away, forming the fainter moving X-ray source well evident in our 2005 \emph{Chandra} observation.

This scenario is of course schematic, and it will require detailed modeling before it can be tested in detail. It is however supported by optical (\citealp{fl94}) and radio (\citealp{rcc+2003}) observations of new blobs appearing repeatedly from behind the obscured region every few years. The basic characteristics of our scenario are based on the detailed study of the simple, homogeneous jets studied by \cite{bop+2006} and rely on the same physics. Proper testing of the scenario (a pulsed jet in a non-homogeneous circumstellar medium) will also require more sensitive X-ray observations, with sufficient statistics as to allow a determination of the spectrum of the moving component (the one plowing through the thinning circumstellar medium), whose temperature is expected to decrease with distance from the star.

\section{Discussion}

The \emph{Chandra} 2005 observation discussed here was designed to provide a clear test of whether the X-ray emission from HH\,154 is due to moving shocks rather than stationary mechanisms such as scattered stellar X-rays. The motion of the X-ray source unambiguously shows that moving shocks play a significant role in the process. However, the observed X-ray morphology is more complex than what would be observed in the presence of a continuously driven jet propagating through a uniform medium. This is not surprising given the evidence from optical observation of a pulsating jet driven from an eruptive process, and the observed non homogeneous distribution of absorbing material in the region surrounding the jet.

Detailed study and modeling of a number of scenarios along the above lines will be needed to determine in detail which conditions can reproduce the observed morphology. While this work is ongoing, simple considerations based on our current models of uniform jets lead us to propose that the most likely scenario explaining the observations is one of an eruptive jet propagating within a medium with strong density gradients. In addition to the (already ongoing) detailed modeling, new, more sensitive X-ray observations will be needed in the future, for example to determine the temperature of the plasma associated with the moving X-ray source, to ascertain whether this is, as expected in such a model, lower than the X-ray temperature associated with the stationary component. These measurement would also provide a direct determination of the scale of density changes in the circumstellar medium around the star, providing an unique probe of such material. Also, they will provide an unique probe of the high energy phenomena at the base of the jet, close to the launch site, constraining the physics of the jet formation.

\acknowledgements

R. B., G. M. and S. S. acknowledge support from MIUR (PRIN 2005-2006) and ASI contract I/023/05/0.
 
\bibliographystyle{aa}

 \end{document}